\documentclass[a4paper,11pt]{article}

\usepackage[dvips]{graphicx}
\usepackage{amsfonts, color, float,  amsmath}
\usepackage{amssymb}
\usepackage{subfigure}
\usepackage{url}
\usepackage{array}
\usepackage{ulem}

\usepackage{hyperref}
\hypersetup{colorlinks=true,
	citecolor=ForestGreen,
anchorcolor=CornflowerBlue,
linkcolor=RoyalBlue,
urlcolor=Cyan,
linkbordercolor={1 1 0}
}

\usepackage{lipsum}  

\usepackage{slashed}
\usepackage[dvipsnames]{xcolor}

\usepackage[small,bf,sf]{caption}  
\usepackage{microtype}
\usepackage{cite}
\usepackage{authblk}
\usepackage[short]{datetime}
\usepackage[bottom]{footmisc}  
\usepackage{booktabs}   

\newdateformat{mydate}{\THEDAY \  \monthname \  \THEYEAR}
\allowdisplaybreaks

\newcommand{\Eq}[1]{Eq.~(\ref{#1})}

\def\beq {\begin{equation}}
\def\eeq {\end{equation}}
\def\bea {\begin{eqnarray}}
\def\eea {\end{eqnarray}}

\def\nn {\nonumber}

\def\bp {\mbox{\boldmath $p$}}

\def\br {\mbox{\boldmath $r$}}

\def\bA {\mbox{\boldmath $A$}}
\def\bB {\mbox{\boldmath $B$}}

\def\bE {\mbox{\boldmath $E$}}
\def\bF {\mbox{\boldmath $F$}}
\def\bff {\mbox{\boldmath $f$}}

\def\bS {\mbox{\boldmath $S$}}
\def\bv {\mbox{\boldmath $v$}}
\def\br {\mbox{\boldmath $r$}}

\def\bj {\mbox{\boldmath $j$}}
\def\bJ {\mbox{\boldmath $J$}}

\def\bn {\mbox{\boldmath $n$}}

\def\brp{\mbox{\boldmath $r$}^{\prime}}
\def\tp{t^\prime}

\def\bnab {\mbox{\boldmath $\nabla$}}

\def\bp {\mbox{\boldmath $p$}}

\def\br {\mbox{\boldmath $r$}}

\def\bA {\mbox{\boldmath $A$}}
\def\bnab {\mbox{\boldmath $\nabla$}}

\AtBeginDocument{}

\title{\LARGE {\bf \sffamily \boldmath On Maxwell's electrodynamics in two spatial dimensions}\vspace{0.3cm}}

\author[]{D.~Boito}
\author[]{L.~N.~S.~de~Andrade}
\author[]{G.~de~Sousa}
\author[]{R.~Gama}
\author[]{C.~Y.~M.~London\vspace{0.3cm}}

\affil[]{\it  Instituto de F\'isica de S\~ao Carlos, Universidade de S\~ao Paulo, CP 369, 13560-970, S\~ao
Carlos, SP, Brazil\vspace{0.3cm}}

\date{}

\begin{document}
\begin{flushright}
{ 
 \mydate \today
}
\end{flushright}

\vspace*{-0.7cm}
\begingroup
\let\newpage\relax
\maketitle
\endgroup

\begin{abstract}
  \noindent
   We discuss the construction of Maxwellian electrodynamics in $2+1$
   dimensions and some of its applications. Special emphasis is
   given to the problem of the retarded potentials and radiation,
   where substantial differences with respect to the usual
   three-dimensional case arise. These stem from the general form of
   the solutions of the wave equation in two dimensions, which we
   discuss using the Green's function method.  We believe the topics
   presented here could be stimulating additions to an advanced
   electrodynamics course at the undergraduate level.

  \end{abstract}

\thispagestyle{empty}

\section{Introduction}
\label{Intro}

Maxwell's electrodynamics is a monumental construction that became one
of the most paradigmatic theories of physics, if not the most
celebrated one.  In undergraduate courses, it is often presented as
the definition of a complete theory: a construction whose internal
machinery alone allows for spectacular predictions, such as the
existence of electromagnetic waves.

In contemporary physics, it has become customary to study successful
physical theories in spacetimes with a different number of dimensions. The
motivation for studies of this type go beyond mere scientific
curiosity.  With the advent of Quantum Electrodynamics (QED) one immediately started
considering its realisations in two and one spatial dimensions.  Schwinger, in 1962, has discussed some
of the peculiarities of QED in a single spatial dimensions~\cite{QED2}--- a theory sometimes referred to as the Schwinger model.  It was shown that quantum corrections generate a mass for the photon without breaking gauge invariance, a rather peculiar fact. Also,  it is known that QED becomes confining in such a limit, i.e. isolated monopoles cannot be found since they would be
confined in neutral bound states and an infinite amount of energy would be required to extract a charge from these ``atoms". Therefore, QED in one spatial dimension is a toy model for Quantum Chromodynamics (QCD), the theory of strong interactions, which does exhibit confinement in our universe. As a matter of fact,  as of today, our understanding of the
spectrum of quark-antiquark bound states is
partially based on a solution of quantum chromodynamics in a
certain limit in only one spatial dimension, due to `t
Hooft~\cite{tHooft}.

It seems, therefore, that classical electrodynamics would be an excellent
framework to introduce undergraduate students to the issues one may
face when studying a physical theory in a spacetime with a different
number of spatial dimensions. However, a search for references on
Maxwell's equations in two spatial dimensions reveals that the
available material is fragmented and scarce. The present work is an
attempt to fill part of this gap in the spirit of an advanced
undergraduate course on electrodynamics. Our main goal is to point out
differences and similarities between Maxwell equations --- and part of
their phenomenology --- in three and two spatial dimensions. We
believe this is an interesting topic to be covered in courses at this
level, and our experience shows that motivated students find very
interesting to understand the construction of this theory and the
potential implications for a hypothetical universe with only two
spatial dimensions.

Part of the material presented here can be found in the literature,
albeit in a fragmented form. About 30 years
ago, a number of papers on the quantum solution of the Hydrogen atom
in $1+1$ and $2+1$ dimensions\footnote{We will refer to the dimensions
  of space-time as $D+1$, where $D$ is the number of spatial
  dimensions.}  were published~\cite{Lapidus,H2D}. In these papers,
one can find short discussions about the electric field and the
Coulomb potential in one and two spatial dimensions, since this is the
only necessary electromagnetic input to the wave-mechanics
calculation.  Classical text books do not discuss the
topic~\cite{Jackson,Griffiths}. To the best of our knowledge, the topic was discussed in the spirit of our work in the Refs.~\cite{Lapidus2,McDonald}.  The unpublished
work of Ref.~\cite{McDonald} by K. T. McDonald is probably the most
self-contained in the literature and it can be considered as the main
reference of the present paper. Here we will borrow part of the
notation and the strategy employed in~\cite{McDonald}. However, we
will develop in detail topics that are not discussed in his work,\footnote{After the online publication of our manuscript, a new version of Ref.~\cite{McDonald} appeared encompassing some of the topics that we
discuss in the present work, as well as correcting an inconsistency found by us in the expressions of the retarded potentials given in the previous version of Ref.~\cite{McDonald}.   } such
as the retarded potentials and radiation in $2+1$ dimensions. We also try
to follow a physical reasoning in order to explain some of the results that can appear as counterintuitive when going
from three to two spatial dimensions. With this aim, it is useful to carefully consider the parallel between the theory in the 2+1 world and usual electrodynamics with one dimension ``stretched out" to infinity, something that was first exploited by Hadamard~\cite{Hadamard}.

In this work, we follow the idea that electrodynamics --- or a theory totally analogous to it --- can exist in any spacetime, regardless of the dimensionality~\cite{Hadamard,McDonald}. In fact, within this assumption, electrodynamics can even be extended to a fractional number of spatial dimensions~\cite{FractalEM}.\footnote{Some authors, however, based on arguments derived from differential geometry prefer to consider that electrodynamics, as we understand it, exists only in spacetimes with odd number of spatial dimensions~\cite{Wheeler}. In these spacetimes, the Faraday tensor and its dual have the same rank, which does not occur in our construction.}
In the construction of electrodynamics in two dimensions one faces a
few issues. The first one, as is well known, is related to the fact
that the Coulomb force must change.\footnote{Even this fact, however,
  has been missed in some of the discussions found in the literature.}
Arguably, the most natural way to construct the theory is to assume that the essence of Gauss' law remains unaltered, but this requires that the
electric field of a point charge now falls off as the inverse of the
distance which, in turn, entails a logarithmic electrostatic
potential. Additionally, part of the vector calculus must change. The
absence of a right-hand rule is obvious and the concept of curl, for
example, must be redefined. As a result of this construction, the magnetic field must be
qualitatively different; it turns out that it cannot be a vector
anymore: it becomes a scalar field. Once
these basic issues are figured out, one can ask how would the multipole expansion
be in such a universe, what are the new retarded potentials, what happens to
electromagnetic waves or, yet, how would dipole radiation
arise. In the remainder of this paper, we discuss
these questions having in mind a curious undergraduate student, one
with particular interest for theoretical physics, or a teacher that
would like to use part of this material in a somewhat advanced course
on electrodynamics.  We will, therefore, stick to usual SI
units. Although sometimes this results in somewhat cumbersome
expressions, it becomes easier to make contact with the usual
undergraduate text books~\cite{Griffiths}.

One of the main differences between electrodynamics in two and three
spatial dimensions appears in the retarded potentials. The reason for
this difference is directly linked to the fact that the solutions to
the wave equation, satisfied by the potentials, change qualitatively
when going from three to two dimensions, a fact that is well known in
the literature in other contexts~\cite{GreensFunc}.  This difference
is often discussed in the framework of Huygens' principle, which
states that every point on a wave front is itself the source of
(spherical) waves. This principle is valid only in spacetimes with
 $D\geq 3$ and odd~\cite{Hadamard}. Huygens' principle relies on
the fact that all waves propagate with a single speed, $c$ in our
case.  In two dimensions, however, a solution to the wave equation can
be understood as a superposition of waves travelling with speeds
ranging from zero to the maximum value $c$, with which the first wave
front travels~\cite{Huygens}.  This fact leads to important differences in the retarded
potentials as compared with the $D=3$ case, and it becomes less obvious how these connect with their
static counterpart. Here, we discuss this issue and the problem of radiation
in the context of a hypothetical universe with 2+1 dimensions. We show that
one benefits from the systematic use of the Green's function method to find the correct form of the retarded potentials. This is sometimes
not emphasized in usual electrodynamics~\cite{Griffiths} where the idea of time retardation is more intuitive, due to the validity of Huygen's principle.

We organize this paper as follows. In Sec.~\ref{Electrostatics}, we discuss
electrostatics in $2+1$ dimensions. The starting point is Gauss' law,
from which we can obtain the electric field and the associated scalar
potencial. In Sec.~\ref{MaxwellEqs}, we will construct the full set of
Maxwell's equations starting from the relativistic formulation. We then discuss
Poynting's theorem, wave equations, the retarded potentials, and electric dipole radiation. The mathematical
notation for vector calculus in $2+1$ dimensions will be introduced
when necessary.  In Sec.~\ref{Conclusions} we summarize the results
and conclude.

\section{Electrostatics in \boldmath \texorpdfstring{$2+1$}{2+1} dimensions}
\label{Electrostatics}

A generalization of Gauss' law is possibly the best starting point to construct electrostatics in 2+1 dimensions. This route is not unique, as we mentioned in the introduction, but is arguably the most natural way to construct the theory. The idea consists in postulating that the essence of Gauss' law is unaltered when
one considers spacetimes with a different number of spatial dimensions: the flux of electric field lines,
$\Phi_E^{(D)}$, through a closed (hyper)surface in $D$ dimensions contains information
about the total charge enclosed by the surface.  For $D=3$, with an electric
charge $q$ at the origin of the coordinate system, and choosing a spherical surface of radius $r$,
this means that
\beq
\Phi_E^{(3)} = |\bE|\, 4\pi r^2 = \frac{q}{\epsilon_0},\label{Gauss3D}
\eeq
from which the usual behavior of the electric field $|\bE| \sim 1/r^2$ can be read off.
Although we are mostly interested in $2+1$ dimensions, it is interesting
to start with a discussion in $D+1$ dimensions.
Considering the Gaussian surface to be a hypersphere of radius $r$, the formulation of Eq.~(\ref{Gauss3D}) can be straightforwardly generalized to
\beq
\Phi_E^{(D)} = |\bE| r^{D-1}\int_S d\Omega_D = |\bE| r^{D-1}  \frac{2\pi^{D/2}}{\Gamma(D/2)} = \frac{q}{\epsilon_0},
\eeq
where we have used the well known result for the integration over the
element of solid angle in $D$ spatial dimensions, $d\Omega_D$, written in terms of the
usual  $\Gamma$ function~\cite{Peskin}. Throughout this work we will use the symbol $\bE$ for the generalized electric field, $q$ for the generalized charge and so on. Their physical meaning follow from usual electrodynamics although there are important differences with respect to their counterpart in 3+1 dimensions, as we discuss below.

Exploiting the hyperspherical symmetry, the electric field of a point charge at the origin can be written as
\beq
\bE = q\frac{\Gamma(D/2)}{2\pi^{D/2}\epsilon_0}\frac{\hat\br}{r^{D-1}}. \label{ED}
\eeq
The dependence of the electric field, and hence of
Coulomb's law, with $r$ is now $1/r^{D-1}$. Thus, the usual 3+1-dimensional fall off as  $1/r^2$
can be understood as a consequence of the dimensionality of our space-time. This point,
albeit somewhat obvious, has sometimes been missed in the literature, as pointed out in Ref.~\cite{Lapidus}.

After this short digression, we can now make $D=2$ in Eq.~(\ref{ED})
to conclude that the electric field $\bE$  of a point charge located at $\brp$ in $2+1$ dimensions
is given by
\beq
\bE(\br) = \frac{q}{2\pi\epsilon_0}\frac{(\br-\brp)}{|\br-\brp|^2},\label{Eq2D}
\eeq
which falls off as the inverse of the distance. One should remark that, as Eq.~(4) shows,  the units of $\epsilon_0$ (and $\mu_0$) change in two dimensions.

The generalization to a
continuous surface charge distribution, $\sigma(\br)$, is direct
\beq
\bE = \frac{1}{2\pi \epsilon_0}\int\! d^2\brp\,\sigma(\brp) \frac{(\br-\brp)}{|\br-\brp|^2}.\label{E2D}
\eeq
Taking the divergence of the last equation\footnote{Remember that now $\bnab = (\partial_x,\partial_y)$.} and using that, for $D=2$,
\beq
\bnab \cdot \left(\frac{\br -\brp}{|\br -\brp|^2}   \right) = 2\pi\, \delta^2(\br -\brp),
\eeq
we recover the differential form of Gauss' law
\beq
\bnab\cdot \bE = \frac{\sigma}{\epsilon_0},
\eeq
which is essentially the same as in the usual space-time with the replacement of the volumetric charge density by a surface one.

Next, we look for an electrostatic potential. In two dimensions, a
vector field $\bF$ is conservative --- and hence can be written in terms of the gradient of a scalar
potential --- if $\partial_x (\bF)_y = \partial_y (\bF)_x$. To make
contact with the idea of the usual curl of a vector, it is useful
to start by defining, in two dimensions, a vector $\bv_\perp$ as
\beq
\bv_\perp = (v_y,-v_x),
\eeq
which is, by construction, perpendicular to the vector $\bv=(v_x,v_y)$.
We define then the operator~\cite{McDonald}
\beq
\bnab_\perp = (\partial_y,-\partial_x).\label{nabperp}
\eeq
With this definition, a vector field is conservative if $\bnab_\perp \cdot \bF =0$. From Eq.~(\ref{nabperp}) and Eq.~(\ref{E2D}) we find that $\bnab_\perp \cdot \bE =0$, and therefore the electrostatic field can be written as $\bE = - \bnab V$. We can then conclude that
the potential obeys, as expected, Poisson's equation in two dimensions
\beq
\nabla^2 V = - \frac{\sigma}{\epsilon_0}.
\eeq
The solution of this equation by the Green's function method yields
\beq
V(\br) = -\frac{1}{2\pi \epsilon_0} \int \! d^2\brp\,\sigma(\brp) \ln \left(\frac{|\br -\brp|}{a}   \right),\label{V2D}
\eeq
from which we can obtain the potential of a point charge at the origin as
\beq
V(r) = -\frac{q}{2\pi \epsilon_0} \ln \left(\frac{r}{a} \right).\label{V2Dpoint}
\eeq
The constant $a$ is an arbitrary reference point that plays a role
similar to that of the additive constant that appears in the potential in  $3+1$ dimensions.
It will also be important to point out that the solution of Eq.~(\ref{V2D}) is completely analogous to the usual three dimensional case, where one has\footnote{For the sake of clarity,  we will often denote three dimensional quantities with an explicit subindex,  as in $V_{3+1}$.}
\beq
V_{3+1}(\br) = \frac{1}{4\pi \epsilon_0} \int \! d^3\brp \frac{\rho(\brp)}{|\br-\brp|},\label{SolPoisson}
\eeq
from which the potential of the point charge at the origin is obtained. 

In the usual electromagnetism, one meets a logarithmic potential of the type of Eq.~(\ref{V2Dpoint})  when studying
the idealised case of an infinite uniformly charged straight wire. Accordingly, the electric field of such a wire
falls off as $1/r$, as in Eq.~(\ref{Eq2D}).
In fact, many of the phenomena of electrodynamics
in two dimensions can be understood as if spacetime were a conveniently chosen two dimensional slice of the usual spacetime with point
charges ``stretched" into infinite wires (an idea pioneered
 by Hadamard~\cite{Hadamard}).
In the case of the field of a point charge, the problem is essentially equivalent to considering the field of the infinite
charged wire in three dimensions on a two-dimensional plane perpendicular to the wire.
In the static case this may seem obvious, but in Sec.~\ref{MaxwellEqs} we will show that even radiation can be understood
in terms of this system where the point charges are extended to infinity.

One should remark that the potential of a point charge in $2+1$ dimensions, Eq~(\ref{V2Dpoint}), is
qualitatively different from the usual $3+1$ case. Note that
Eq.~(\ref{V2Dpoint}) diverges when $r\to \infty$ (in addition to the
divergence at the origin, which is also present in $3+1)$. This means
that a hydrogen atom in $2+1$ dimensions would be a {\it confined}
system: an infinite amount of energy would be needed to extract the
electron from the atom, in direct analogy to what happens
in quark-antiquark bound states where, at large distances,  a linear potential, usually written as $V(r)\approx \sigma r$, confines the particles inside the bound state~\cite{Cornell}. Therefore, in a universe with $2+1$
dimensions chemistry would be radically different~\cite{H2D}.
In addition, the total energy of a point charge is infinite not only due to the divergence for small distances, as in the usual 3+1 case, but also for large distances as well. In fact,  any charge distribution with non-zero net charge will carry an infinite amount of energy. This very peculiar fact is in part solved by the hypothesis of confinement: i.e. if charges in such a theory are confined, as color charge is confined in QCD, then all charge distributions would have a zero monopole term and isolated charges would not be observed in the universe.

In the construction of electrostatics, the discontinuity of the electric field plays an important role.
In 3+1 dimensions, the electrostatic field is discontinuous when
crossing a surface charge. Accordingly, in 2+1, the discontinuity appears when
crossing a line charge. The boundary conditions can be
found with the usual method, by considering a Gaussian ``box'', which now
becomes a rectangle in two dimensions, around the line charge
$\lambda$. Making the lateral size of the box approach zero one finds
that the component of the electric field perpendicular to the surface is
discontinuous by an amount given by \beq
\label{eq:condcampo}
\bE_{\rm above} - \bE_{\rm below} = \dfrac{\lambda}{\epsilon_0}\hat \bn,
\eeq
where $\bE_{\rm above}$  and $\bE_{\rm below}$ are defined by the choice of the normal vector $\hat \bn$.

The scalar potential is continuous (as in 3+1 dimensions), but again the gradient of the potential inherits the discontinuity of the electric field
\beq
\bnab V_{\rm above} - \bnab V_{\rm below} = -\dfrac{\lambda}{\epsilon_0} \hat{\bn}.
\eeq
The last two equations are examples of aspects of the construction that are completely straightforward, and follow exactly what one could intuitively expect, with the surface charge replaced by the line charge $\lambda$.

One can proceed developing the standard electrostatics tool box in 2+1
dimensions. Let us comment, for example, on the multipole
expansion of the scalar potential for points away from the source. The result of Eq.~(\ref{V2D}) can be
straightforwardly expanded for $r > r^\prime$. In polar coordinates
one finds
\beq
\label{Vmulti}
V(r,\phi) = - \dfrac{Q_{\rm T}}{2 \pi \epsilon_0} \ln \left(\frac{r}{a}\right) + \dfrac{1}{2 \pi \epsilon_0} \hspace{0.1cm} \sum \limits _{n=1}^{\infty} \, \dfrac{A_n \cos (n\phi) + B_n \sin (n\phi)}{n \, r^n},
\eeq
with the coefficients $A_n$ and $B_n$ given by
\beq
A_n = \int d^2 \brp (r^\prime)^n\sigma(\brp)  \cos (n\phi') \hspace{0.1cm} , \qquad  B_n = \int d^2 \brp (r^\prime)^n\sigma(\brp) \sin (n\phi') \hspace{0.1cm}.
\eeq
The structure of the expansions shows that we can still talk about a ``monopole term",  a ``dipole term", and so on, although their behaviour with $r$ is now different compared with usual electrodynamics. The monopole term contains the total charge $Q_{\rm T}$  and is, as expected, logarithmic. Charge confinement would essentially forbid the monopole term in two dimensions and therefore $Q_T=0$. It is followed by a dipole term, which would be the leading effect here that
falls off as $1/r$, then by a quadrupole term that behaves as $1/r^2$ and so
on. By considering only the dipole term one can then write
\beq
V_{\rm dip}(\br)= \frac{1}{2\pi \epsilon_0\, r}\int d^2 \brp\, r^\prime\, \sigma(\br^\prime) \cos(\phi-\phi^\prime),
\eeq
which can be cast as
\beq
V_{\rm dip}(\br) =  \frac{1}{2\pi \epsilon_0} \frac{ \hat\br\cdot \bp}{r},
\eeq
with the dipole vector moment given by
\beq
\bp =  \int d^2\brp\, \brp \,  \sigma(\br^\prime).
\eeq
One clearly sees that in the multipole expansion, apart from the  different power counting in $1/r$, the only qualitatively different term is the leading monopole, which becomes logarithmic in 2+1 dimensions.

The general solution to the Poisson equation in two dimensions using
the method of separation of variables can be found in almost any book
on mathematical methods for physics. With Eq.~(\ref{Vmulti}) it is
easy to make contact with these results and establish the standard
correspondence between $V(\br)$ obtained from the multipole expansion
and Poisson's equation in polar coordinates.  Typical electrostatic
problems can be solved with these techniques and their solution do not
differ significantly from the familiar ones in 3+1 dimensions. In the
remainder, we will focus on electrodynamics and the problem of radiation,
where more subtle issues arise.

\section{Electrodynamics and radiation}
\label{MaxwellEqs}

An elegant strategy to obtain the full set of Maxwell's equations in
2+1 dimensions consists in using as a starting point the Faraday
tensor, $F_{\mu\nu}$. For completeness, and to establish the notation, we
briefly review in our conventions Maxwell's equations in 3+1 dimensions
in the manifestly covariant form. We define the four-potential  $A^\mu = \left( V/c, \bA \right)$ formed
from the scalar potential $V$ and the vector potencial $\bA$.\footnote{We use $g_{\mu\nu} ={\rm diag}(+1,-1,-1,-1)$. We will keep all factors of $c$ explicit and, accordingly, we work with the coordinate $x^0=ct$ and $\frac{\partial}{\partial x^0} =\frac{\partial}{c \partial t}$.}  We then build the tensor $F^{\mu\nu} = \partial^\mu A^\nu - \partial^\nu A^\mu$  which explicitly reads, in our conventions,
\begin{equation}\label{eq:F3D}
F_{3+1}^{\mu \nu} =
\begin{pmatrix}
0 & -E^x/c & -E^y/c & -E^z/c \\
E^x/c & 0 & -B^z & B^y \\
E^y/c & B^z & 0 & -B^x \\
E^z/c & -B^y & B^x & 0
\end{pmatrix}.
\end{equation}
Forming the dual tensor $\tilde F_{3+1}^{\mu\nu} =\epsilon^{\mu\nu\sigma\rho} F_{\sigma\rho}$, Maxwell's equations are then written as
\bea
&&\partial_\mu F_{3+1}^{\mu \nu} = \mu_0 J^\nu, \label{Max3D1}  \\
&& \partial_\mu \tilde{F}_{3+1}^{\mu \nu} = 0. \label{Max3D2}
\eea
where $J^\mu$ is the four-current defined as $J^\mu = \left( c \rho, \bJ \right)$ with $\rho$ and $\bJ$ being respectively the volume charge and current densities.

In 2+1 dimensions, Faraday's tensor has now only three independent components. These will contain  $E_x$ and $E_y$ as well as (the now scalar) magnetic field $B$. In a world where all electric effects are confined to the $xy$ plane, the magnetic field would be along the $z$ direction. This is sufficient to see that the last row and the last column of $F^{\mu\nu}$ given in Eq.~(\ref{eq:F3D}) must now absent.   Faraday's tensor in 2+1 dimensions can then be obtained by ``removing'' the last row and the last column of Eq.~(\ref{eq:F3D}).\footnote{The same procedure can be applied once more to obtain $F^{\mu\nu}$ in 1+1 dimensions, as discussed in more detail in~\cite{McDonald}. We relegate to appendix \ref{Fmunu} an alternative derivation of this result.}  It then reads~\cite{Lapidus2,McDonald}
\begin{equation}\label{eq:F_2D}
F^{\mu \nu} =
\begin{pmatrix}
0 & -E^x/c & -E^y/c \\
E^x/c & 0 & -B \\
E^y/c & B & 0 \\
\end{pmatrix}.
\end{equation}
The immediate consequence is that now the magnetic field has only its would-be $z$ component; it is no longer a vector field and it becomes a (pseudo) scalar field. The equivalent
of Eq.~(\ref{Max3D1}) can be obtained in a straightforward manner. To write the 2+1 equivalent of Eq.~(\ref{Max3D2}) one must notice that now the dual tensor will become a vector, since the Levi-Civita symbol loses one index. We have then $\tilde F^{\mu} = \epsilon^{\mu\nu\sigma}F_{\nu\sigma}$. In covariant form, Maxwell's equations in 2+1 dimensions are then
\bea\label{eq:Max_2D}
&&\partial_\nu F^{\mu \nu} = \mu_0 j^\mu, \\
&&\partial_\nu \tilde{F}^{\nu} = 0,
\eea
where in $j^\mu$ the surface charge and current densities appear.

Diffentiating Eq.~(\ref{eq:Max_2D}) and remembering that the electromagnetic tensor is antisymmetric, we obtain
\beq
\partial_\mu\,j^\mu=0,
\eeq
which is the covariant form of the continuity equation. As in three dimensions, electric charge is conserved.

One can then
work out the differential Maxwell's equations in 2+1 dimensions in terms of the fields as
\bea
&&\bnab \cdot \bE = \frac{\sigma}{\epsilon_0},    \\
&&\bnab_\perp \cdot \bE = \frac{\partial B}{\partial t},   \label{Maxwell2} \\
&&\bnab_\perp B = \mu_0 \bj + \frac{1}{c^2}\frac{\partial \bE}{\partial t},\label{Maxwell3}
\eea
where $\sigma$ and $\bj$ are the surface charge and surface current,
respectively, and we have made use, again, of the operator of
Eq.~(\ref{nabperp}). We have also used that in 2+1 it
remains true that $c^2 = (\mu_0\epsilon_0)^{-1}$ (as we will show
below).
 One notices that there is no equivalent to the three-dimensional magnetic Gauss' law.

Electrodynamics becomes complete with the Lorentz force law. From the expression of the Lorentz force in terms of the Faraday tensor and the 4-velocity, $u_\mu$, which reads $f^\mu=(q/c)F^{\mu\nu}u_\nu$, one obtains in 2+1 dimensions the following result
\beq
\bff=q(\bE+\bv_\perp \, B).
\eeq

An immediate consequence of Maxwell's equations in 2+1 dimensions is the existence of electromagnetic waves in two spatial dimensions. To follow the usual steps done in the
derivation of the wave equations for $\bE$ and $B$ one needs the following two-dimensional vector relation $\bnab_{\perp}(\bnab_{\perp} \cdot \bF)= \bnab^2\bF  - \bnab(\bnab\cdot\bF)$. Using it in \Eq{Maxwell2} we get $\bnab_{\perp}(\bnab_{\perp} \cdot \bE) = \partial_t(\bnab_{\perp}B)$ and consequently
\begin{equation}\label{eq:wave_E}
	\nabla^2\bE - \frac{1}{c^2} \partial^2_{tt}\bE = \frac{1}{\epsilon_0} \bnab \sigma + \mu_o \partial_t \bj,
\end{equation}
which is the general wave equation with sources. A similar procedure gives the wave equation for the magnetic field, which is now a scalar wave
\begin{equation}\label{eq:wave_B}
	\nabla^2 B - \frac{1}{c^2} \partial^2_{tt} B = \mu_0 \bnab_{\perp} \cdot \bj.
\end{equation}

In the context of electromagnetic waves, a crucial role is played by
the Poynting vector.  It may not be obvious how to construct this vector in two
spatial dimensions. In order to closely follow the familiar derivation, it turns out to be useful to work with
the field $\bE_\perp = (E_y,-E_x)$. One can then follow the familiar steps
of the derivation of energy balance in the electromagnetic field. Starting from the power $\frac{dW}{dt} = \int_A d^2\brp \left( \bE \cdot \bj \right)$,
and replacing $\bj$ by its expression in terms of fields using Amp\`ere-Maxwell's law, \Eq{Maxwell3}, and using the vector identity  \[
\bnab \cdot(\bE_{\perp}B) = B(\bnab_{\perp} \cdot \bE) + \bE \cdot (\bnab_\perp B)
\]
one finally finds
\begin{equation*}
\frac{dW}{dt} = -\frac{d}{d t}\int_A d^2\brp \left[ \frac{B^2}{2\mu_0} + \frac{\epsilon_0 E^2}{2} \right] - \frac{1}{\mu_0} \int_A d^2\brp ~\bnab \cdot \left( \bE_\perp B \right).
\end{equation*}
Upon use of the divergence theorem, one can then relate the derivative with respect to time of the mechanical and electromagnetic energies to the flux  of
the Poynting vector, defined as
\beq
\bS = \frac{1}{\mu_0} \bE_\perp B\label{Poynting}
\eeq
through a closed loop. Or, in the usual differential form,
\beq
\frac{\partial}{\partial t} \left( u_{\text{mec}} + u_{\text{em}} \right) + \bnab \cdot \bS = 0.
\eeq
The meaning of the Poynting vector given in Eq.~(\ref{Poynting}) can be understood as follows.
 The vector $\bE_\perp B$ is perpendicular to $\bE$ and by construction points in a direction that would also be perpendicular to the would-be third dimension. In other words, if we consider in 3+1 dimensions the fields $\bE=(E_x,E_y,0)$ and $\bB=(0,0,B)$ the Poynting vector would be proportional to $\bE \times \bB = (E_y\,B,-E_x\,B,0)$, which has the same $x$ and $y$ components as $\bE_\perp B$ in 2+1 dimensions.

The requirement for radiating fields is that they carry energy away
from the system to infinity.  For this to happen, since now the flux
of the Poynting vector is to be computed through a closed loop, one
obtains that for radiating fields $\bS \sim 1/r$, suggesting that
 radiation fields should fall off as $\bE_{\rm rad} \sim 1/\sqrt{r}$ and $B_{\rm rad}
\sim 1/\sqrt{r}$.

As in 3+1 dimensions, it is advantageous to treat the problem of
radiation starting from the potentials $V$ and $\bA$.
To follow a similar route in $D=2$, we must first establish the relation between the vector potential and the magnetic field. This can be done in a way consistent with Eq.~(\ref{eq:F_2D}) through~\cite{McDonald}
\beq
B = -\bnab_{\perp} \cdot \bA.
\eeq
From Faraday's law one concludes that $\bnab_{\perp} \cdot (\bE + \partial_t\bA) = 0$ and, thus,
 the quantity $\bE + \partial_t\bA$ is conservative and can be written as the gradient of a scalar potential, which implies  $\bE =  -\bnab V - \partial_t\bA$. As usual,
gauge invariance allows one to choose $\bnab \cdot \bA = -\frac{1}{c^2} \partial_t V$, which corresponds to Lorenz gauge.\footnote{L. Lorenz, Danish physicist, not to be confused with H. Lorentz, the Dutch physicist famous for the Lorentz transformations~\cite{Whittaker}.} The equations satisfied by the potentials are then the wave equation with sources. In 2+1 dimensions, and reminding the reader that $\nabla^2 = \partial_x^2 +\partial_y^2$, the equations are\footnote{To facilitate the comparison with the usual expressions, the potentials and Green's functions will always be written as functions of the variable $t$ (instead of $x^0=ct)$.}
\bea
  \nabla^2 V- \frac{1}{c^2} \partial^2_{t t} V &=& -\frac{\sigma}{\epsilon_0} , \label{eq:wave_V}\\
  \nabla^2 \bA - \frac{1}{c^2}\partial^2_{tt} \bA&=& -\mu_0 \bj.\label{eq:wave_A}
  \eea

  At this point, we must look for the solutions of
  Eqs.~(\ref{eq:wave_V}) and (\ref{eq:wave_A}). In the usual 3+1 case,
  they would be the standard retarded potentials. Here, an important
  difference between three and two spatial dimensions occurs. In three
  dimensions, the retarded potentials can be obtained in a rather
  intuitive, albeit not very rigorous way, by simply imposing in the
  static solution that in the dynamical case information travels with
  the speed of light, as suggested by the wave equations satisfied by
  $V$ and $\bA$. In practice, one is allowed to take the solution to
  Poisson's equation, Eq.~(\ref{SolPoisson}), and simply replace in
  the integrand $\rho(\brp) \rightarrow \rho(\brp, t_{r^\prime})$ with
  the retarded time given by $t_{r^\prime}=t-|\br -\brp|/c$. This
  solution is correct in three dimensions and has the advantage of
  being very intuitive.

  A more formal derivation of the retarded potentials
  can be obtained using the Green's function method, which will be the basis for our
  discussion of the two-dimensional case. It is, therefore, useful to briefly discuss this solution in $D=3$. We define the
  Green's function of the d'Alembertian operator in three dimensions as~\cite{Appel}
  \beq
\left(\nabla^2 - \frac{1}{c^2} \partial^2_{tt}\right)G_{3+1}(\br,t) \equiv \square^2 G_{3+1}(\br,t) = - 4\pi \delta^3(\br)\delta(t) .\label{GreenDef3D}
\eeq
 The solution for $V(\br,t)$ is obtained, by construction, as
\beq
V_{3+1}(\br, t) = \frac{1}{4\pi \epsilon_0} \int d^3\brp \int dt^\prime G_{3+1}(\br-\brp,t-t^\prime)\, \rho(\brp, t^\prime).\label{eq:V3Dgen}
\eeq
 In 3+1 dimensions, the Green's function of the d'Alembertian is, explicitly,
\beq
G_{3+1}(\br-\brp,t-\tp) = \frac{\delta\left(\tp - t_{r^\prime}\right)}{|\br -\brp|}  .\label{G3D}
\eeq
The integration over $t^\prime$ immediately leads to the usual form of
the retarded potentials. For the scalar potential we have
\beq
V_{3+1}(\br, t) = \frac{1}{4\pi \epsilon_0} \int d^3\brp \, \frac{\rho(\brp,t_{r^\prime})}{|\br-\brp|}. \label{Vret3D}
\eeq

The Green's function of Eq.~(\ref{G3D}) can be
interpreted as the potential associated with a point charge that exists at
point $\brp$ in space only at the instant $\tp$, as per Eq.~(\ref{GreenDef3D}).\footnote{This is, of course, inconsistent with charge conservation, but it remains useful as a mathematical device to interpret the Green's function. The same applies to the non-physical charge distribution of Eq.~(\ref{eq:rhowiret}).\label{fn:chargeconservation}} The Dirac $\delta$ in the
numerator shows that at another point in space, $\br$, at a later time, $t$,
the effect of the existence of this point charge that blinks at $t=\tp$
will only be felt once, precisely at the time $t= \tp + |\br-\brp|/c$.
In turn, this is linked to the fact that the general solution to the spherical
wave equation in three dimensions is given by functions of the type $f(r\pm ct)/r$, i.e., waves that propagate with a single speed $c$.
This is sometimes called ``sharp'' wave propagation, and is a feature
that is consistent with Huygens' principle~\cite{Appel}.

Let us now turn to our case of interest, namely the retarded
potentials in two spatial dimensions. Now, a naive attempt to obtain the
solution by simply replacing $\sigma(\brp) \rightarrow \sigma(\brp,
t_{r^\prime})$ in Eq.~({\ref{V2D}}) is wrong. This is so
because the intuitive idea about the way information propagates, embedded in this naive solution,  breaks down in two spatial dimensions --- in other words, Huygens' principle is not valid in two spatial dimensions~\cite{Huygens}. To understand
what is happening it is better to use the Green's function method.
Now we have
\beq
\square^2 G_{2+1}(\br,t)= -2\pi \delta^{(2)}(\br) \delta (t).
\eeq
The solution for this Green's function can be found in the literature~\cite{GreensFunc}, it reads
\beq
G_{2+1}(\br-\brp,t-\tp) =  \frac{\theta(c\tau-s)}{\sqrt{\tau^2-s^2/c^2}},  \quad \mbox{with} \quad s = \left| \br - \brp \right|, \quad \tau = t - t^\prime.\label{G2D}
\eeq
Instead of the Dirac delta we now have a Heaviside $\theta$
function. This function is still zero for $t<\tp +
|\br-\brp|/c$, but the effects of the ``point charge'' that existed at
$t=\tp$ are felt {\it forever} after this time. There is, however, an attenuation given by the denominator of Eq.~(\ref{G2D}) and the effect disappears, asymptotically, with time. The outermost wave front propagates with velocity $c$, in agreement with the expectations of causality but, as we discuss in detail below, the full wave can be written as a superposition of modes with velocities ranging from zero to $c$, which is
not a sharp wave propagation as in three dimensions~\cite{Appel}.\footnote{A possible interpretation for this ``tail" of the Green's function is put forward in Ref.~\cite{Dai:2013cwa} in terms of a charge moving with the speed of light at the light-cone shell.}

The analogy with the three dimensional infinite charged wire is again instrumental in understanding this effect.  Consider, in three dimensions,  an infinite charged wire along the $z$ axis with constant charge density  $\lambda=2\pi\eta$ (where $\eta$ is a constant that carries the dimensions) that exists only at instant $t=0$, with charge distribution given by
\beq
\rho(\br,t)=2\pi\eta\,\delta(x)\delta(y)\delta(t).\label{eq:rhowiret}
\eeq
We can then obtain its potential on a two dimensional slice of three dimensional space, which we take to be the $xy$ plane.
 With Eqs.~(\ref{G3D}) and~(\ref{Vret3D}), for a point at distance $r=\sqrt{x^2+y^2}$ from the $z$ axis we have
\beq
 \frac{ \epsilon_0V(\br,t)}{\eta} = \int_{-\infty}^{\infty} dz \frac{1}{2 } \frac{\delta(t - \sqrt{z^2+r^2}/c)}{\sqrt{z^2+r^2}}.\label{Vwire}
\eeq
Information from two points with $z\equiv \pm\,\bar z=\pm\sqrt{(ct)^2-r^2}$ arrive simultaneously at a point on the $xy$ plane at time $t$.
The Dirac $\delta$ in the integrand can then be written as
\beq
 \delta(t - \sqrt{z^2+r^2}/c) =\frac{ct}{\sqrt{t^2-r^2/c^2}} \left[ \delta(z - \sqrt{c^2 t^2 - r^2}) + \delta(z + \sqrt{c^2 t^2 - r^2}) \right].
\eeq
Clearly, if $ct<r$ the integral will give zero. Using this fact and the $\delta$ functions to perform the integral we find
\beq
 \frac{ \epsilon_0V(\br,t)}{\eta}= \frac{\theta(ct - r)}{\sqrt{t^2-r^2/c^2}},
\eeq
which is identical to the two-dimensional Green's function of Eq.~(\ref{G2D}), and therefore Eq.~(\ref{Vwire}) is an integral representation of Eq.~(\ref{G2D}). In this example, it becomes clear that information from two equidistant points
at $z=\pm \bar z$ will be reaching a given point on the $xy$ plane for $t=r/c$ and since information travels with the speed of light from every point of the infinite wire, the potential is felt forever.
 For the observer confined to the two-dimensional slice of spacetime, information would appear to emanate from the origin
 but travelling with different speeds $v(z) = \frac{rc}{\sqrt{z^2+r^2}}$, ranging from 0 to $c$.

 The last observation provides us with some additional intuition regarding the Green's function $G_{2+1}(r,t)$ given in Eq.~(\ref{G2D}) and suggests the following practical result.  From the integral representation we derived above we can write
 \beq
 G_{2+1}(r,t) = \int_{0}^{\infty} dz  \frac{\delta(t - \sqrt{z^2+r^2}/c)}{\sqrt{z^2+r^2}},\label{eq:G2Int}
 \eeq
and apply the change of variables $v(z) = \frac{rc}{\sqrt{z^2+r^2}}$ suggested by the analogy with the infinite charged wire to obtain
\beq
G_{2+1}(r,t) = \int_{0}^{c} dv  \frac{c/v}{\sqrt{c^2 - v^2}} \delta(t - r/v),
\eeq
 in agreement with the interpretation given above, namely that information in two dimensions appears to travel with different speeds, ranging from 0 to $c$, which immediately implies that Huygen's principle is not valid.

Having discussed the Green's function in two dimensions in detail, one can then write the solution for the retarded potentials. The analog to Eq.~(\ref{eq:V3Dgen}) in 2+1 dimensions~is
\beq
V_{2+1}(\br, t) = \frac{1}{2\pi \epsilon_0} \int d^2\brp \int dt^\prime G_{2+1}(\br-\brp,t-t^\prime)\, \sigma(\brp, t^\prime),\label{eq:V2Dgen}
\eeq
 and the solutions to $V(\br)$ and $\bA(\br)$ can be written as\footnote{A superseded version of Ref.~\cite{McDonald} contained a mistake in the expression of the retarded potentials. The present version of Ref.~\cite{McDonald}, published after our work was made public, is now correct and in full agreement with our expressions.}
\bea
V(\br) &=& \frac{1}{2\pi \epsilon_0} \int d^2\brp \int dt^\prime ~\frac{\theta(c\tau-s)}{\sqrt{\tau^2-s^2/c^2}} \sigma(\brp, t^\prime)  , \label{retV2D}     \\
     \bA(\br) &=& \frac{1}{2\pi \epsilon_0} \int d^2\brp\int dt^\prime ~\frac{\theta(c\tau-s)}{\sqrt{\tau^2-s^2/c^2}} \bj(\brp, t^\prime),
\eea
with the notation of Eq.~(\ref{G2D}).
In two dimensions it is, therefore, less trivial to make contact with the static case, given for the scalar potential in Eq.~(\ref{V2D}). In particular, it is not as obvious how the potentials of a point charge appear in the static limit. It is interesting then to consider an unphysical case which violates charge conservation, but that
is instrumental in understanding how the new formulae for the retarded potentials are compatible with the static ones. Let us take the potentials for an oscillating monopole\footnote{This choice is partially motivated by the fact that we will use this oscillating monopole as a building block for the physical dipole when studying radiation. Another possible route is the direct use of the integral representation of the Green's function in Eq.~(\ref{eq:G2Int}).}
with localized charge distribution at an arbitrary point $\br_0$ given by
\beq
\sigma(\br, t) = q \delta^2(\br -\br_0) e^{i \omega t}.
\eeq
(One is tacitly assuming that the physical charge is connected to the real part of the last expression.)
The static limit can be recovered doing $\omega\to 0$.
Using this charge density in the scalar potential the integrals
over $\brp$ can be done with the delta function present in $\sigma(\br,t)$ and we find
\beq
V(\br,t) = \frac{q}{2\pi \epsilon_0} \int\limits_{-\infty}^{+\infty} d\tp \frac{\theta(c\tau-d)}{\sqrt{\tau^2-d^2/c^2}}e^{i\omega \tp} = \frac{q}{2\pi \epsilon_0} \int\limits_{-\infty}^{t-d/c} d\tp \frac{e^{i\omega \tp}}{\sqrt{(t-\tp)^2 - d^2/c^2}},
\eeq
where we used the $\theta$ function and $d=|\br-\br_0|$ is the distance from $\br$ to the charge at $\br_0$.
With a change of variables, this last integral can be cast as
\beq
V(\br,t) =  \frac{q e^{i \omega t}}{2 \pi \epsilon_0} \int_{1}^{\infty} \mathrm{d}z ~\frac{e^{-i \omega z d/c}}{\sqrt{z^2 - 1}}.
\eeq
In this form, it becomes simple to identify that the solution is a combination of Bessel functions of the first and second kind, known as the Hankel function of the second kind,  $H_\nu^{(2)}(z)$, defined in Eq.~(\ref{H02}) in the appendix. Using the relevant integral representations of Eq.~(\ref{Bessel}), we explicitly obtain
\beq
V(\br,t) =  -\frac{q e^{i \omega t}}{4 \epsilon_0} i H_0^{(2)} \left(\frac{\omega |\br -\br_0|}{c} \right). \label{monopole2D}
\eeq
In the static limit, $\omega \to 0$, and using that  $H_0^{(2)} \left( z \rightarrow 0 \right) \sim -\frac{2i}{\pi} \ln z$, we recover precisely the potential
of a point charge located at $\br_0$ as
\beq
V(\br,t)= -\frac{q}{2\pi \epsilon_0} \ln \left(\frac{|\br -\br_0|}{a}\right),
\eeq
which shows that Eq.~(\ref{retV2D}) is perfectly compatible with the static case.

Let us now turn to the problem of dipole radiation. We can
model an oscillating electric dipole in two dimensions adapting the usual
procedure followed in three dimensions. We consider two tiny disks at points $\br_+$ and $\br_-$ connected by a thin wire that carries
a current back and forth from one to the other, such that if one disk has
charge $q(t)$ the other has charge $-q(t)$ and the system is always neutral.
More precisely, the charge distribution of such a dipole can be written as
\beq
\sigma_{\rm dip}(\br, t) = q e^{i\omega t} \delta^{(2)}(\br-\br_+) - q e^{i\omega t} \delta^{(2)}(\br-\br_-).
\eeq
Formally, it is the superposition of two oscillating monopoles and the associated exact scalar potential $V(\br,t)$  can
be obtained from the solution Eq.~(\ref{monopole2D})
\beq
V_{\rm dip}(\br,t)=-\frac{q e^{i\omega t}}{4\epsilon_0} i \left[ H_0^{(2)} \left( \frac{\omega}{c} \left| \br - \br_+ \right| \right) - H_0^{(2)} \left( \frac{\omega}{c} \left| \br - \br_- \right| \right) \right].
\eeq
Here we are interested in the radiation zone, where $r\gg r_{\pm}$. We can then use $r_{\pm}/r\ll 1$ and Taylor expand the Hankel functions to find
\beq
V_{\rm dip}(\br,t)\approx-\frac{ e^{i\omega t}}{4\epsilon_0} i \left[   -\frac{\omega}{c} \hat\br \cdot \bp \left. \frac{d H_0^{(2)}(z)}{d z} \right|_{z = \omega r / c}  +\cdots \right]
\eeq
where we used without loss of generality that $\br_+=-\br_-$ and the dipole moment is $\bp=2q\br_+$. As expected, the monopole term cancels and we are left with a dipole term, which contains the part of $V(\br,t)$ that will contribute to radiating fields.  We now use the asymptotic form of $H_0^{(2)}(z)$ for $z\to \infty$ of Eq. (\ref{HankelAsymp}) to obtain
\beq
V_{\rm dip}(\br,t)\approx -\frac{\hat{\br} \cdot \bp}{2 \epsilon_0} \sqrt{\frac{\omega}{2 \pi c}} \frac{e^{i(\omega (t - r/c) + \pi/4)}}{\sqrt{r}} +\cdots\label{VdipRad}
\eeq
This expression is the part of the scalar potential that contributes to dipole radiation since its gradient will generate terms in $\bE$ that behave as
$1/\sqrt{r}$ as required for radiating fields in two spatial dimensions.
To make contact with the familiar expressions in three dimensions, one
can consider an electric dipole at the origin with $\bp$ aligned along the  $\hat y$ direction.
In this case,  taking the real part of Eq.~(\ref{VdipRad}),  one finds
\beq
V_{\rm dip}(\br,t)\approx -\frac{qr_+ \cos(\tilde\phi)}{2\pi \epsilon_0} \sqrt{\frac{2\omega\pi}{  c}} \frac{\sin[\omega (t - r/c) + \delta_0]}{\sqrt{r}}   +\cdots \label{FinalResult}
\eeq
where $\tilde \phi$ is the angle between $\hat\br$ and the $y$-axis and $\delta_0$ is a constant phase. In the final result for the potential in the radiation zone, the usual retarded time with respect to the origin intervene.
 It is interesting to note that, although it is crucial in the intermediate steps to consider the somewhat counterintuitive solutions
 of the wave equation in 2+1 dimensions, the behaviour of $V_{\rm dip}(\br,t)$ in the radiation zone is essentially the same found in three dimensions, and the outgoing electromagnetic wave travels with the speed of light.

\section{Conclusions}
\label{Conclusions}

We have discussed the construction of Electrodynamics in two spatial
dimensions. We tried to emphasise mainly the somewhat unexpected
differences that arise with respect to the usual three-dimensional
case. First, we began with a general discussion of Gauss' law and how
the Coulomb field changes with the dimensions of space-time. It is
well known that Coulomb's force, or Newton's law of gravitation
attraction, is expected to behave as the inverse of the distance (to
the first power) in two spatial dimensions. This leads to a
logarithmic potential which, in itself, is enough to dramatically
alter the phenomenology of electromagnetism in 2+1 dimensions, since
the attractive potential between opposite charges becomes confining,
i.e., an infinite amount of energy would be required to extract the
electron from the hydrogen atom, for example. It is not difficult to
obtain the general solution of Poisson's equation in this context and
to consider its multipole expansion. This expansion retains its main
features, in spite of a logarithmic monopole term and the different
power counting in $1/r$ with the dipole term behaving as $1/r$ and so
on.

Next, we considered the construction of the full set of Maxwell's
equations. One must face the fact that the magnetic field becomes a
scalar field. With a few definitions for the vector calculus in two
dimensions, one is able to construct Maxwell's equations in a form
that is similar to the usual one, although there is no equivalent for
the magnetic Gauss' law. With these equations at hand, it becomes
straightforward to find the wave equations satisfied by electric and
magnetic fields and by the scalar and vector potentials (in Lorenz
gauge). Here, an important difference with respect to the three
dimensional case appears. Huygen's principle is not valid in spaces
with even number of dimensions. In practice, this means that part of
our intuition regarding the way information propagates is no longer
valid since information travels as a superposition of waves with
speeds ranging from 0 to $c$. The retarded potentials in two
dimensions have a rather different form, inherited from the Green's
function of the d'Alembertian in 2+1 dimensions which has a Heaviside
theta function instead of the usual Dirac delta. We have then shown
that this solution, albeit less intuitive, is fully consistent with
the static case.

We then turned to an investigation of radiation, using the
paradigmatic oscillating dipole. We were able to show in
Eq.~(\ref{FinalResult}) that, in spite of the qualitative difference
in the behaviour of the solutions of the wave equation in two and
three dimensions, the scalar potential in the radiation zone retains a
form very similar to the one we are used to in three
dimensions. Therefore, dipole radiation in two dimensions is much more
similar to its counterpart in three dimensions than the form of the
retarded potentials may suggest.

We should state very clearly that we do not know of any physical system,
e.g. a material, where electrodynamics would be described by the
two-dimensional Maxwell's equations we discussed here.  Rather, our
main purpose with this note was to collect a few results on
electrodynamics in two dimensions that could be useful for an advanced
course on electromagnetism at the undergraduate level.  However, in
this spirit, other interesting aspects of the phenomenology of
Maxwell's theory in 2+1 dimensions remain to be investigated such as
the radiation of a magnetic dipole, to name one.  We believe the
issues we discussed here are, nevertheless, sufficient to show that
the subject is interesting, and can be the basis, for example, for
student's projects. These would certainly broaden their knowledge of
electromagnetism and introduce them to some of the common practices of
theoretical physics.  Our experience confirms this fact, since the
results presented here were, in part, obtained as a project by four of
us while undergraduate students.

\section*{Acknowledgements}
We would like to thank Leonardo Maia, and Diogo Soares Pinto for
stimulating discussions. We thank K.~T.~McDonald for bringing Refs.~\cite{FractalEM,Wheeler} to our attention and
Daniel Vanzella for insightful discussions about the derivation presented in App.~B and for bringing to our attention Refs.~\cite{Wald,Vanzella}.
 This work has received partial support from
the S\~ao Paulo Research Foundation (FAPESP grant  \#2015/20689-9) and from CNPq
(305431/2015-3).

\appendix

\section{Useful Bessel function properties}

\renewcommand{\theequation}{\thesection.\arabic{equation}}
\setcounter{equation}{0}

We list here a few results from Abramowitz and Stegun~\cite{AS} that are useful when studying solutions to the wave equation in two spatial dimensions.
The following integral representations for the Bessel functions of the first and second kind were used
\bea
J_\nu(x) &=& \frac{2(\frac{1}{2}x)^{-\nu}}{\pi^{1/2} \Gamma\left( \frac{1}{2} - \nu\right)} \int_{1}^{\infty} dt \frac{\sin (xt) }{\left( t^2 - 1 \right)^{\nu + 1/2}}, \nn
    \\
    Y_\nu(x)& =& -\frac{2(\frac{1}{2}x)^{-\nu}}{\pi^{1/2} \Gamma\left( \frac{1}{2} - \nu\right)} \int_{1}^{\infty}dt  \frac{\cos (xt) }{\left( t^2 - 1 \right)^{\nu + 1/2}}.\label{Bessel}
\eea
The Hankel functions of the second kind are defined as
\beq
H_\alpha^{(2)}(x) = J_\alpha(x) -iY_\alpha(x).\label{H02}
\eeq
It is also useful to know the following asymptotic forms of $H_0^{(2)}(z)$ for $z\to 0$  and $z\to \infty$
\bea
 &&H_0^{(2)} \left( z \rightarrow 0 \right) \approx -\frac{2i}{\pi} \ln z, \\
 &&H_0^{(2)}(z \to \infty) \approx \sqrt{\frac{2}{\pi z}} e^{-i(z - \pi/4)}\label{HankelAsymp}.
\eea

\section{\boldmath \texorpdfstring{$F^{\mu\nu}$}{Fmunu} in \texorpdfstring{$2+1$}{2+1} dimensions}
\label{Fmunu}
We present here an alternative derivation of our result for $F^{\mu\nu}$ in 2+1 dimensions.
In the usual 3+1 theory, Faraday's tensor can be decomposed as
\begin{equation}
  F^{\mu \nu} = \frac{1}{c^2} \left( E^\mu u^\nu-  E^\nu u^\mu   \right) - \frac{1}{c}\epsilon^{\mu \nu \delta \gamma } u_\delta B_\gamma,
\end{equation}
where we used $\epsilon_{0123}=+1$ and the fact that an observer moving with four-velocity $u^\mu$ interprets $E^\mu = F^{\mu\nu}u_\nu$ as the electric
field and $B^\mu= \frac{1}{2c}\epsilon^{\mu\nu\sigma\rho}F_{\sigma\rho}u_\nu$ as the magnetic field~\cite{Wald} (see also Ref.~\cite{Vanzella} for a related discussion).  From this expression, the usual 3+1 components of $F^{\mu\nu}$, Eq.~(\ref{eq:F3D}), can be obtained for the case of an observer at rest with $u^\mu=(c,0,0,0)$.

In 2+1 dimensions the Levi-Civita tensor loses one index and the magnetic field is no longer a vector which immediately leads to
\beq
F^{\mu \nu} = \frac{1}{c^2} \left(E^\mu u^\nu-  E^\nu u^\mu  \right) -  \frac{1}{c}\epsilon^{\mu \nu \gamma } u_\gamma B.
\eeq
Specifying to the observer at rest, one then derives our Eq.~(\ref{eq:F_2D}).


\begin{thebibliography}{99}

\bibitem{QED2} J.~S.~Schwinger,
  {\it ``Gauge Invariance and Mass. 2.,''}
  Phys.\ Rev.\  {\bf 128}, 2425 (1962).

\bibitem{tHooft} G.~'t Hooft, {\it A Two-Dimensional Model for Mesons,}
  Nucl.\ Phys.\ B {\bf 75}, 461 (1974).


\bibitem{Lapidus} I.~Richard~Lapidus, {\it One- and Two-Dimensional Hydrogen Atoms}, Am. J. Phys. {\bf 49}, 807 (1981).


\bibitem{H2D}  F.~J.~Asturias and S.~R.~Arag\'on, {\it The hydrogenic atom and the period table of the elements
  in two spatial dimensions}, Am. J. Phys. {\bf 53}, 893 (1985).


\bibitem{Jackson} J.~D.~Jackson, {\it Classical electrodynamics}, (Wiley, New York, 1975).

\bibitem{Griffiths} D.~J.~Griffiths {\it Introduction to Electrodynamics}, (Prentice Hall, Upper Saddle River, NJ, 1999).

\bibitem{Lapidus2}  I.~Richard~Lapidus, {\it  Classical electrodynamics in a universe with two space dimensions}, Am. J. Phys. {\bf 50}, 155 (1982).

\bibitem{McDonald} K.~T.~McDonald, {\it Electrodynamics in 1 and 2 spatial dimensions}, \url{http://www.physics.princeton.edu/~mcdonald/examples/2dem.pdf}.


\bibitem{Hadamard} J.~Hadamard, {\it Lectures on Cauchy's Problem}, (Yale Univ.\ Press, New Haven, 1923).

\bibitem{FractalEM} M. Zubair, M. J. Mughal, Q. A. Naqvi, {\it Electromagnetic Fields and Waves in Fractional Dimensional Space}, (Springer, Heidelberg, 2012).

\bibitem{Wheeler}N. Wheeler, {\it ``Electrodynamics" in 2-Dimensional Spacetime},
\url{https://www.reed.edu/physics/faculty/wheeler/documents/Electrodynamics/Miscellaneous\%20Essays/E&M\%20in\%202\%20Dimensions.pdf}

  \bibitem{GreensFunc} K. Watanabe, {\it Integral Transform Techniques for Green's Function,} Lecture Notes in Applied and Computational Mechanics 76, Springer International Publishing Switzerland (2014).



\bibitem{Huygens} K. Brown, {\it Huygens principle}, \url{https://www.mathpages.com/home/kmath242/kmath242.htm}.


\bibitem{Peskin} M.~E.~Peskin and D.~V.~Schroeder,
  {\it An Introduction to quantum field theory,} Reading, USA: Addison-Wesley (1995).


\bibitem{Cornell} E.~Eichten, K.~Gottfried, T.~Kinoshita, K.~D.~Lane and T.~M.~Yan,
  {\it Charmonium: The Model,}
  Phys.\ Rev.\ D {\bf 17}, 3090 (1978)
  Erratum: [Phys.\ Rev.\ D {\bf 21}, 313 (1980)].


\bibitem{Whittaker} E. T. Whittaker, {\it A history of theories of aether and electricity}, (Humanities Press, New York, 1973), p. 267.

  \bibitem{Appel} W.~Appel, {\it Mathematics for Physics and Physicists}, Princeton: Princeton University Press, (2007).


  \bibitem{Dai:2013cwa}
  D.~C.~Dai and D.~Stojkovic,
  {\it Origin of the tail in Green's functions in odd-dimensional space-times},
  Eur.\ Phys.\ J.\ Plus {\bf 128}, 122 (2013),
  [arXiv:1309.2996 [hep-th]].


  \bibitem{AS} M.~Abramowitz and I.~A.~Stegun,  {\it Handbook of Mathematical Functions with Formulas, Graphs, and Mathematical Tables.} (Dover, New York, 1964).

  \bibitem{Wald} R. M. Wald, {\it General Relativity}, (Chicago, The University of Chicago Press, 1984),  p.~64.

\bibitem{Vanzella} D. A. T. Vanzella, G. E. A. Matsas, H. W. Crater, {\it Comment on ``General covariance, Lorentz covariance, the Lorentz force, and Maxwell equations,"}, Am. J. Phys. 62 (10), 923 (1994).

\end{thebibliography}
\end{document}